\documentclass[preprint2]{aastex}

\newcommand{\secname}{Section}
\newcommand{\secsname}{Sections}
\newcommand{\secref}[1]{\secname~\ref{#1}}
\newcommand{\secrefii}[2]{\secsname~\ref{#1} and~\ref{#2}}
\newcommand{\eqn}{Equation}
\newcommand{\eqns}{Equations}
\newcommand{\eqnref}[1]{\eqn~\ref{#1}}
\newcommand{\eqnrefii}[2]{\eqns~\ref{#1} and~\ref{#2}}
\newcommand{\fig}{Figure}
\newcommand{\figs}{Figures}
\newcommand{\figref}[1]{\fig~\ref{#1}}
\newcommand{\figrefii}[2]{\figs~\ref{#1} and~\ref{#2}}
\newcommand{\tab}{Table}
\newcommand{\tabref}[1]{\tab~\ref{#1}}

\newcommand{\thetitle}{Minimal prospects for radio detection of extensive air showers in the atmosphere of Jupiter}

\shorttitle{Radio detection of extensive air showers in the atmosphere of Jupiter}
\shortauthors{Bray \& Nelles}

\begin{document}
\newcommand{\aipcs}{AIP Conf.\ Series}   % American Institute of Physics
\newcommand{\arep}{Astron.\ Rep.}        % Astronomy Reports
\newcommand{\arxiv}{ArXiv e-prints}      % ArXiv
\newcommand{\app}{Astropart.\ Phys.}     % Astroparticle Physics
\newcommand{\cpc}{Comput.\ Phys.\ Commun.} % Computer Physics Communications
\newcommand{\cphysc}{Chinese Phys.\ C}   % Chinese Physics C
\newcommand{\geoji}{Geophys.\ J.\ Int'l} % Geophysical Journal International
\newcommand{\nima}{Nucl.\ Instrum.\ Meth.\ A} % Nuclear Instruments & Methods in Physics Research Section A
\newcommand{\njp}{New J.\ Phys.}         % New Journal of Physics Conference Series
\newcommand{\plb}{Phys.\ Lett.~B}        % Physics Letters B
\newcommand{\privcom}{priv.\ comm.}      % private communication
\newcommand{\procsci}{Proc.\ Sci.}       % Proceedings of Science
\newcommand{\radres}{Radiation Res.}     % Radiation Research
\newcommand{\rsla}{Proc.\ R.\ Soc.\ A}   % Proceedings of the Royal Society of London Series A: Mathematical, Physical and Engineering Sciences
\newcommand{\sci}{Science}               % Science
\newcommand{\spjetp}{Sov.\ Phys.\ JETP}  % Soviet Journal of Experimental and Theoretical Physics

\title{\thetitle}

\author{J.\ D.\ Bray}
\affil{JBCA, School of Physics \& Astronomy, University of Manchester, Manchester M13 9PL, UK}
\email{justin.bray@manchester.ac.uk}

\author{A.\ Nelles}
\affil{Department of Physics \& Astronomy, University of California, Irvine, CA 92697, USA}

\begin{abstract} % not more than 250 words
 One possible approach for detecting ultra-high-energy cosmic rays and neutrinos is to search for radio emission from extensive air showers created when they interact in the atmosphere of Jupiter, effectively utilizing Jupiter as a particle detector.  We investigate the potential of this approach.  For searches with current or planned radio telescopes we find that the effective area for detection of cosmic rays is substantial (\mbox{$\sim 3 \times 10^7$}~km$^2$), but the acceptance angle is so small that the typical geometric aperture (\mbox{$\sim 10^3$}~km$^2$\,sr) is less than that of existing terrestrial detectors, and cosmic rays also cannot be detected below an extremely high threshold energy (\mbox{$\sim 10^{23}$}~eV).  The geometric aperture for neutrinos is slightly larger, and greater sensitivity can be achieved with a radio detector on a Jupiter-orbiting satellite, but in neither case is this sufficient to constitute a practical detection technique.  Exploitation of the large surface area of Jupiter for detecting ultra-high-energy particles remains a long-term prospect that will require a different technique, such as orbital fluorescence detection.
\end{abstract}


\section{Introduction}

The spectrum of ultra-high-energy cosmic rays decreases steeply at energies above $10^{20}$~eV \citep{abraham2010}.  Detection of cosmic rays around and above this threshold, and potentially neutrinos at similar energies, may help to clarify whether this cut-off is due to interactions of propagating cosmic rays or an inherent limit in the spectra of their sources, to locate the positions of these sources on the sky \citep{abraham2007}, and to determine whether there is a contribution to the cosmic-ray flux from exotic top-down mechanisms \citep[e.g.][]{aloisio2015}.  However, detecting the rarer particles at higher energies requires detectors with extremely large apertures.

Such large apertures may potentially be obtained through remote monitoring of planet-sized bodies \citep{gorham2004b} such as Earth \citep[e.g.][]{takahashi2009} or the Moon \citep[e.g.][]{bray2014a}.  As the largest body available in the solar system --- apart from the Sun --- Jupiter is an attractive option for this approach, and the detection of cosmic-ray interactions in its atmosphere through their gamma-ray or radio emission has been proposed \citep{rimmer2014} and debated \citep{privitera2014}.

Codes are available to simulate the extensive air shower\footnote{Although ``air'' strictly refers to the atmosphere of Earth, we use the term ``air shower'' to refer to a particle cascade in any atmosphere.} produced when an ultra-high-energy cosmic ray or neutrino interacts in an atmosphere \citep{sciutto1999}, and the resulting radio emission \citep{alvarez-muniz2012c}.  These codes have been validated against terrestrial air showers, and can be extended to the case of Jupiter.  In this work we use these codes to analyze the development (\secref{sec:development}) and radio emission (\secref{sec:emission}) of a Jovian air shower.  In \secrefii{sec:cosrays}{sec:neutrinos} we examine the interaction geometry and the consequent geometric aperture for cosmic rays and neutrinos respectively, and in \secref{sec:prospects} we consider the prospects for detecting the radio emission from these with a realistic experiment.  We discuss in \secref{sec:disc} the implications of our results for the potential of Jupiter as an ultra-high-energy particle detector, and briefly summarise our conclusions in \secref{sec:conc}.

\section{Development of Jovian air showers}
\label{sec:development}

When a cosmic ray interacts in the atmosphere of Jupiter, as on Earth, a shower of particles develops, generating and entraining additional particles until the energy per particle drops low enough that ionization losses dominate, and the shower dissipates.  The environment on Jupiter differs from Earth in several ways which will affect the development of a shower.
\begin{itemize}
 \item The atmosphere of Jupiter is 75\% molecular hydrogen by mass, with the remainder composed almost entirely of helium.  Several characteristic quantities are therefore quite different, as shown in \tabref{tab:params}.  In particular, the radiation length is almost doubled.
 \item Due to the greater scale height of the Jovian atmosphere, a shower penetrating down through a given column density will develop in a less dense medium, as for showers on Earth with a large inclination angle relative to the vertical.  As we shall see in \secref{sec:cosrays}, the geometry for a detectable Jovian cosmic-ray air shower requires that it be highly inclined, exacerbating this effect.
 \item The magnetic field of Jupiter is stronger than that of Earth, with a strength of 400~$\mu$T at the equator and 1100--1400~$\mu$T at the poles \citep{smith1974}, versus equivalent values of 35~$\mu$T and 65~$\mu$T respectively for the geomagnetic field \citep{finlay2010}.
\end{itemize}
To determine the effect of this differing environment on the development of an air shower, we carry out a series of simulations with the AIRES code \citep{sciutto1999}.  AIRES has been developed to simulate showers in the terrestrial atmosphere, so to represent the Jovian atmosphere we use the TIERRAS extension \citep{tueros2010}, which allows for simulations in other media such as ice or soil.  Using this extension we define a custom Jovian atmosphere, using the values for a hydrogen-helium mixture from \tabref{tab:params} (except for the density, which we vary between simulations).  For simplicity, in all simulations we take the primary particle to be a cosmic-ray proton.  The resulting shower may differ from that produced by a neutrino, or by the heavier cosmic-ray nuclei at higher energies \citep{aab2014}, but the radio emission is primarily determined by the energy deposited in the dominant electromagnetic component of the shower \citep{nelles2015}, so a proton-initiated shower is an adequate model for any case with a similar shower energy.

% However, the radio emission is mostly dominated by the electromagnetic component of the shower, which is similar for all hadronic primaries, and it has been shown on Earth that the radio emission of a shower depends primarily on its geometry and not on the nuclear composition of its primary \citep{nelles2015}.  Since the shower geometry does not vary between hadronic primaries with respect to the the distance of the observer (Jupiter-Earth), any hadronic primary is sufficient to calculate the radio emission. 

\begin{deluxetable}{crcccc}
 \tablewidth{0pt}
 \tablecaption{Standard parameters for Jovian and terrestrial atmospheric gases \label{tab:params}}
 \tablehead{
  \colhead{Composition} &
  \colhead{Density\tablenotemark{a}} &
  \colhead{Refractive index\tablenotemark{a}} &
  \colhead{Radiation length} &
  \colhead{Effective $Z$} &
  \colhead{Mean $Z/A$}
  \\
  \colhead{(by mass)} &
  \colhead{(g/cm$^{3}$)} &
  \colhead{} &
  \colhead{(g/cm$^{2}$)} &
  \colhead{} &
  \colhead{}
 }
 \startdata
  100\%~H$_2$
   & $7.1 \times 10^{-5}$ % 0.000071
   & 1.000132
   & 63.04
   & 1\phd\phn\phn\phn
   & 0.9921 \\
  100\%~He
   & $12.5 \times 10^{-5}$ % 0.000125
   & 1.000035
   & 94.32
   & 2\phd\phn\phn\phn
   & 0.4997 \\
  75\%~H$_2$/25\%~He\tablenotemark{b}
   & $8.5 \times 10^{-5}$ % 0.000085
   & 1.000118
   & 68.74
   & 1.257\tablenotemark{c}
   & 0.8690 \\
  terrestrial air
   & $120.5 \times 10^{-5}$ % 0.001205
   & 1.000289
   & 37.10
   & 7.265
   & 0.4992
 \enddata
 \tablenotetext{a}{At standard temperature and pressure.}
 \tablenotetext{b}{Representative of the Jovian atmosphere.}
 \tablenotetext{c}{Calculated per method~II of \citet{henriksen1957}.}
 \tablerefs{\citep{olive2014,sciutto1999}}
\end{deluxetable}

% Note: the values above are mostly taken or straightforwardly derived from olive2014; except for effective Z values for mixtures, which are derived in a more complex fashion, using a formula from henriksen1957, using sciutto1999 for the composition for terrestrial air.  Values for terrestrial atmosphere are the Linsley standard atmosphere used by AIRES; they're not specified in the AIRES paper (sciutto1999), but they are in the associated code, which will have to do.

\begin{figure}
 \plotone{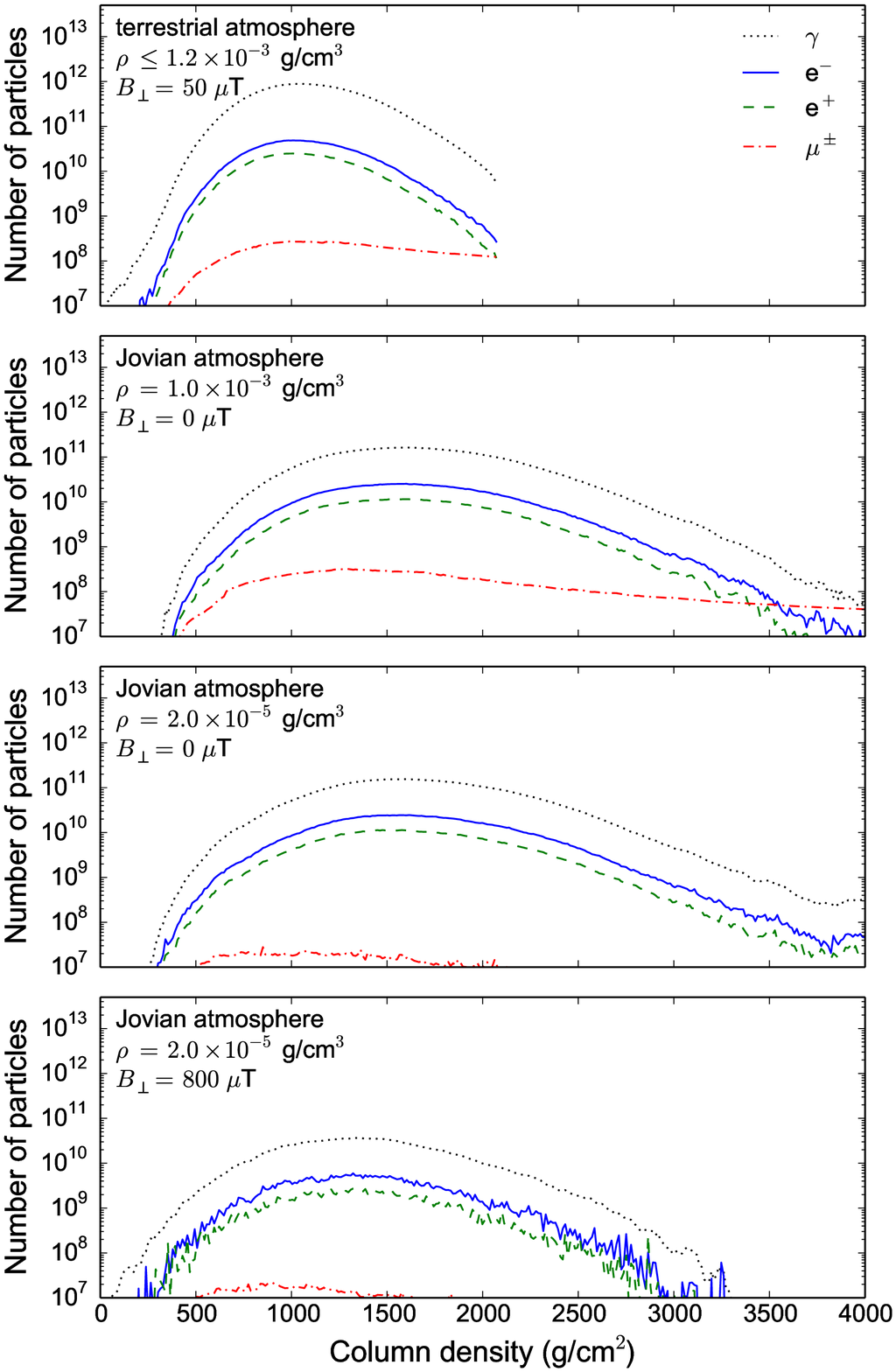}
 \caption{Longitudinal development of simulated air showers in terrestrial and Jovian conditions, showing the dominant particle species.  The primary particle in each case is a cosmic-ray proton with an energy of $10^{20}$~eV.  Compared to the terrestrial air shower (top), a shower in the Jovian atmosphere with comparable density $\rho$ (upper center) develops at a greater column density because of the increased radiation length of this medium.  In the less-dense upper atmosphere of Jupiter (lower center), where a detectable cosmic-ray air shower is more likely to develop (see \secref{sec:cosrays}), muons are less numerous, because the larger physical distance associated with a given column density makes them more likely to decay.  The introduction of the Jovian magnetic field with a realistic transverse field strength $B_\perp$ (bottom) causes the shower to be strongly suppressed by synchrotron losses.}
 \label{fig:particles}
\end{figure}

The longitudinal development profiles of some simulated showers are shown in \figref{fig:particles}, isolating the effects of each of the environmental differences listed above.  The composition of the Jovian atmosphere causes the electromagnetic cascade to be elongated compared to a terrestrial shower, the reduced density suppresses the muon flux by allowing more decays, and the Jovian magnetic field, using a representative strength of 800~$\mu$T, causes the electromagnetic cascade to be strongly suppressed, as the high-energy electrons and positrons rapidly lose energy to synchrotron radiation.  The synchrotron photons in this regime are high-energy gamma rays which continue to participate in the shower, but this loss still causes the shower to initially develop and attenuate more rapidly than when electrons and positrons produce gamma rays primarily through bremsstrahlung as in a classical electromagnetic cascade.  This last effect, in particular, reduces the peak number of charged particles by a factor \mbox{$\sim 4$}, significantly decreasing the potential for detecting the radio emission of a Jovian air shower.

\section{Radio emission from Jovian air showers}
\label{sec:emission}

The positively and negatively charged particles in an extensive air shower, primarily electrons and positrons, are deflected in opposite directions by the local magnetic field.  As electrons and positrons are continuously produced and deflected over the life of the shower, they give rise to a time-varying transverse current which is responsible for the majority of the shower radio emission.  Additional emission arises from the time-varying excess of negative charge in the shower \citep{askaryan1962}, although this effect is usually small for showers in gaseous media.  The beam pattern of the emission depends on the refractive index of the atmosphere, being enhanced at the Cherenkov angle.  The basic principles of the emission mechanisms have been understood for some time \citep{kahn1966,allan1971}, and modern microscopic simulations produce results that closely match radio observations of terrestrial air showers; see \citet{huege2016} for a review.

In this work we extend our simulations from \secref{sec:development} by calculating the radio emission with the ZHAireS code \citep{alvarez-muniz2012c}, which has been validated against observations of terrestrial air showers \citep[e.g.][]{buitink2016}.  An additional parameter required by this code is the refractive index of the Jovian atmosphere, which we take to be
 \begin{equation}
  n_{\rm r} = 1 + 0.000014 \times \left( \frac{\rho}{10^{-5}~{\rm g/cm}^{3}} \right)
 \end{equation}
in terms of the density $\rho$, for non-standard temperature and pressure, based on the values in \tabref{tab:params}.

\begin{figure}
 \plotone{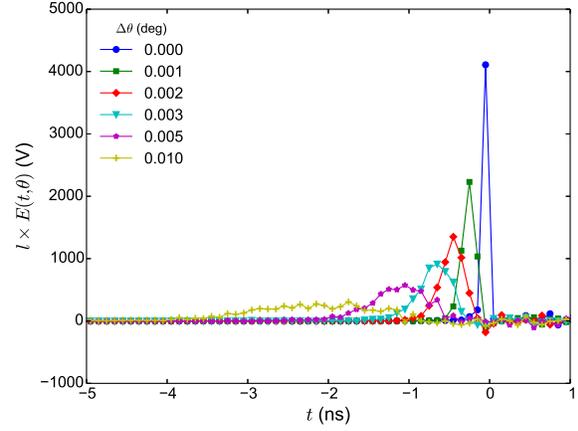}
 \caption{Radio pulses from a simulated Jovian air shower, showing electric field strength $E(t, \theta)$ at distance $l$ in the far field.  At greater angles $\Delta\theta$ inside the Cherenkov cone, the pulse is broader and weaker; angles outside the Cherenkov cone (not shown) display the same effect.  The shower shown here had a primary cosmic-ray proton with an energy of $10^{20}$~eV, and developed in a Jovian atmosphere with a density of \mbox{$2 \times 10^{-5}$}~g/cm$^{-3}$.}
 \label{fig:sim_pulse}
\end{figure}

For an observer at the Cherenkov angle
 \begin{equation}
  \theta_{\rm c} = \arccos(1/n_{\rm r})
 \end{equation}
from the shower axis, the entire shower is observed near-simultaneously.  Coherence over the length of the shower, and hence the amplitude of the emitted radio pulse, are therefore maximized when the observer is within a small angle $\Delta\theta$ from the Cherenkov angle.  Sample pulses are shown in \figref{fig:sim_pulse}, illustrating this.
% The polarization shown in this figure and hereafter is the polarization of the peak of the pulse. % Probably don't need to bother saying this.

\begin{figure}
 \plotone{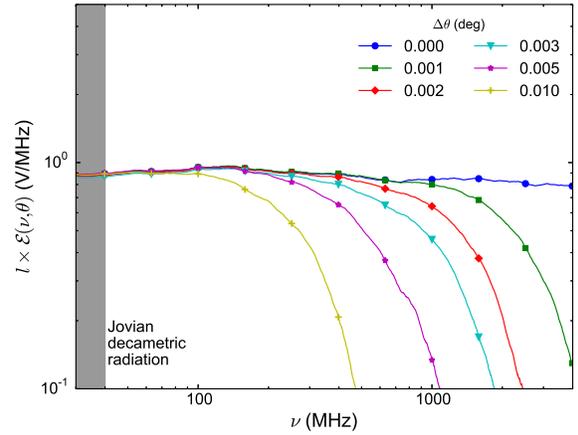}
 \caption{Spectra of the simulated radio pulses shown in \figref{fig:sim_pulse}, showing spectral electric field strength $\mathcal{E}(\nu, \theta)$, smoothed as described in the text.  At greater angles $\Delta\theta$ inside the Cherenkov cone, the pulse loses coherence and becomes weaker at higher frequencies.  The shaded region shows the frequency range in which detection of the pulse is impractical because of the background Jovian decametric radiation.}
 \label{fig:sim_spect}
\end{figure}

To find the spectra of these pulses over frequency $\nu$, we take their complex Fourier transform, convolve this with a variable-width window of fractional bandwidth \mbox{$\Delta\nu/\nu = 0.5$}, and take the magnitude of the result.  This represents an observation with a corresponding experimental bandwidth, but the effect of varying this assumption is minor: for practical purposes, this is just a smoothing operation.  The resulting spectra are shown in \figref{fig:sim_spect}.  Depending on $\Delta\theta$, these spectra generally cut off at frequencies below a few~GHz.  At low frequencies, observations are limited by the background Jovian decametric radiation, with an intensity of \mbox{$\sim 10^6$}~Jy \citep{warwick1967}.  This radiation cuts off sharply at a frequency of 40~MHz, corresponding to the maximum cyclotron frequency in the local magnetic field, so this may be taken as a lower limit to the frequency for practical detection of Jovian air showers.

\begin{figure}
 \plotone{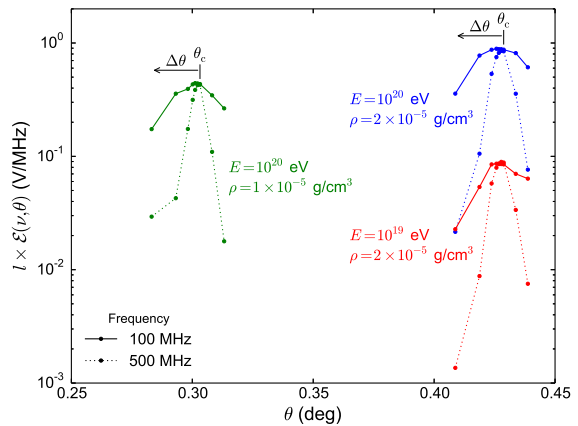}
 \caption{Beam patterns of radio pulses from simulated Jovian air showers, showing spectral electric field strength $\mathcal{E}(\nu, \theta)$ at distance $l$ in the far field.  The radiation is beamed as a hollow cone at the Cherenkov angle \mbox{$\theta = \theta_{\rm c}$} from the axis of the shower.  If the shower develops at a lower atmospheric density (left; green), then the emission is weaker than for a similar shower at a higher density (top-right; blue), and the Cherenkov angle is smaller.  If the shower has a lower energy (bottom-right; red), the emission is weaker but the Cherenkov angle is unchanged.  In all cases, the radiation has a smaller beam width scale $\Delta\theta$ at higher frequencies.}
 \label{fig:sim_beam}
\end{figure}

The beam pattern of the emission around the Cherenkov angle, and the dependence on the energy $E$ of the shower and the density of the medium, are shown in \figref{fig:sim_beam}.  Taking a frequency-dependent Gaussian beam shape per \citet{alvarez-muniz2006}, and assuming a linear dependence on shower energy and a power-law dependence on density, we apply a rough parameterization of the spectral electric field as
 \begin{eqnarray}
  \mathcal{E}(\nu,\theta;E,\rho,l) =
   \mathcal{E}_0
   \left( \frac{E}{10^{20}~{\rm eV}} \right)
   \left( \frac{\rho}{10^{-5}~{\rm g/cm}^3} \right)^{\!\!\alpha}
   \nonumber \\ \times
   \left( \frac{l}{8 \times 10^{11}~{\rm m}} \right)^{\!\!-1}
   \exp\!\left(
    -\left( \frac{\Delta\theta}{\theta_{\rm c}} \right)^{\!\!2}
    \left( \frac{\nu}{\nu_0} \right)^{\!\!2}
   \right)
   \label{eqn:parameterisation}
 \end{eqnarray}
where \mbox{$l = 8 \times 10^{11}$}~m is the mean distance from Earth to Jupiter.  For the free parameters in this expression we fit the values
 \begin{eqnarray}
  \mathcal{E}_0 &=& 5.7 \times 10^{-7}~\mu{\rm V/m/MHz} \\
  \nu_0 &=& 6.9~{\rm MHz} \\
  \alpha &=& 0.94
 \end{eqnarray}
using a series of simulations spanning the ranges $10^{19}$~eV to $10^{21}$~eV in energy and \mbox{$1 \times 10^{-5}$}~g/cm$^{3}$ to \mbox{$5 \times 10^{-5}$}~g/cm$^{3}$ in density.  To test the accuracy of these values, we simulate a single additional shower, and compare its radio emission with that predicted by the parameterization.  We find that the parameterization predicts the width of the emission beam in the angular variable $\Delta\theta$ to within a factor of two, which will determine the resulting error in the geometric apertures calculated in \secrefii{sec:cosrays}{sec:neutrinos}.  We expect the linear dependence on $E$ in \eqnref{eqn:parameterisation} to hold reasonably well outside the energy range used for our fit, as this linearity is a consistent feature of coherent pulses from particle cascades.

\section{Geometric aperture for cosmic rays}
\label{sec:cosrays}

\begin{figure}
 \plotone{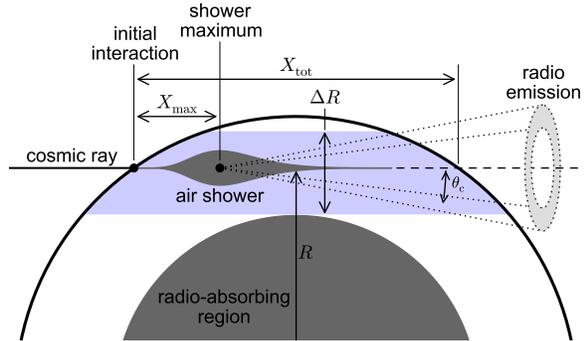}
 \caption{Interaction geometry for a detectable Jovian air shower.  The shower develops from the initial interaction, reaching its maximum development at a column depth \mbox{$X_{\rm max} \sim 1500$}~g/cm$^{2}$.  The radio emission from the shower is beamed forward as a hollow cone at the Cherenkov angle $\theta_{\rm c}$ around the projected shower axis (dashed).  For the shower to be detectable, it must be sufficiently deep that the total column density \mbox{$X_{\rm tot} > X_{\rm max}$} allows it to completely develop, but not so deep that the radio emission is absorbed.}
 \label{fig:schematic}
\end{figure}

For a cosmic ray interacting in the Jovian atmosphere to be detected, it must meet two conditions: the interaction must be sufficiently deep that the extensive air shower will fully develop and produce coherent radio emission, and sufficiently shallow that this radio emission --- directed forward along the axis of the shower --- is not completely attenuated as it escapes the atmosphere.  These conditions constrain detectable cosmic rays to trajectories that skim the atmosphere of Jupiter, as described by \citet{rimmer2014} and illustrated in \figref{fig:schematic}.  The detection region is an annulus around Jupiter with a width of $\Delta R$ and a circumference of $2 \pi R_{\rm J}$, where \mbox{$R_{\rm J} = 6.9 \times 10^4$}~km is the mean radius of Jupiter.

\begin{figure}
 \plotone{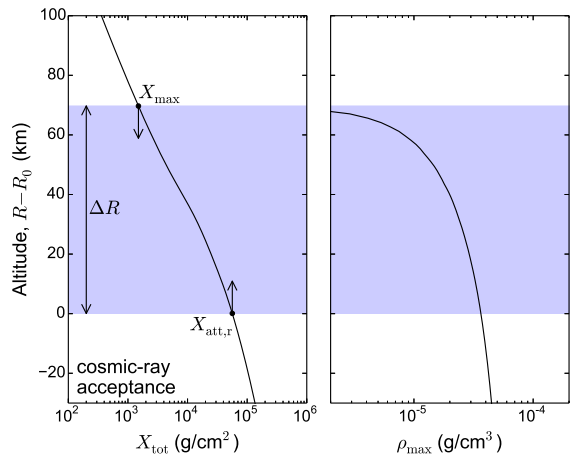}
 \caption{Range of altitudes (shaded) for the lowest point on the projected path of a detectable cosmic ray through the Jovian atmosphere, at radius $R$ as illustrated in \figref{fig:schematic}.  As shown in the left panel, we require that the column density $X_{\rm tot}$ along the path of the cosmic ray be at least \mbox{$X_{\rm max} \sim 1500$}~g/cm$^{2}$, and no greater than the radio attenuation length $X_{\rm att,r}$.  The right panel shows $\rho_{\rm max}$, the density at a column depth of $X_{\rm max}$, at which the shower development is maximized.  The reference zero altitude, at radius $R_0$, is at a pressure of 1000~mbar.}
 \label{fig:atmo}
\end{figure}

For the full development of an air shower, we require that the total column density $X_{\rm tot}$ of the Jovian atmosphere along its path is sufficient for it to reach its point of maximum development, at a column depth of \mbox{$X_{\rm max} \sim 1500$}~g/cm$^{2}$.  We calculate $X_{\rm tot}$ from the atmospheric density profile of \citet{moses2005}, similarly to \citet{rimmer2014}, and find the altitude of the lowest point on the axis of the air shower for which \mbox{$X_{\rm tot} > X_{\rm max}$}, as shown in \figref{fig:atmo}.  This is not a precise limit, as showers above this altitude will have non-zero radio emission, and showers slightly below this altitude will have reduced radio emission due to their truncation soon after their point of maximum development.  However, due to the exponential profile of the atmosphere, modifying this assumption will have a relatively small effect on the maximum altitude $R_{\rm max}$ of the detection region.

For the radio emission to escape, we require that the projected shower axis pass no deeper than a pressure of 1000~mbar, at which \citet{lindal1981} found the $S$-band signal from the Voyager~1 probe at 2.3~GHz to be extinguished due to absorption by ammonia.  This condition implies that the radio emission is assumed to be unattenuated up to a radio attenuation length of \mbox{$X_{\rm att,r} = 6 \times 10^4$}~g/cm$^{2}$, calculated from the model of \citet{moses2005} as above, and completely attenuated beyond this threshold.  This is around an order of magnitude larger than the attenuation length we calculate from measurements with the Galileo probe \citep{folkner1998}, and from the model of \citet{janssen2005}, but these results are for near-vertical transects, and are therefore more sensitive to conditions at low altitudes, with a greater relative concentration of radio-absorbing ammonia.  As the Jupiter-skimming geometry of the Voyager measurement better represents the air-shower geometry considered in this work, we will use the larger number for the radio attenuation length, and acknowledge that it may lead to an optimistic value for the cosmic-ray aperture.  We also neglect the frequency dependence of the radio attenuation; but, as for $R_{\rm max}$ above, the exponential profile of the atmosphere will restrict these approximations to a relatively small effect on the minimum altitude $R_{\rm min}$ of the detection region.

Between the altitudes $R_{\rm min}$ and $R_{\rm max}$ we find a range of \mbox{$\Delta R = 70$}~km as shown in \figref{fig:atmo}.  The resulting area of the annular detection region is \mbox{$2 \pi R_{\rm J} \Delta R = 3 \times 10^7$}~km$^2$, which is substantial: it exceeds the area of the Pierre Auger Observatory \citep{aab2015b}, the largest current cosmic-ray detector, by four orders of magnitude.  However, a cosmic ray passing through this area will only be detected if its beamed emission is directed towards a radio antenna.  The resulting acceptance angle for a detectable cosmic ray, assuming a single radio antenna (e.g.\ at Earth), has the same solid angle as the emission, or
 \begin{equation}
  \Omega = 4 \pi \, \theta_{\rm c} \, \Delta\theta_{\rm max}
  \label{eqn:solidang}
 \end{equation}
where $\Delta\theta_{\rm max}$ is the maximum separation from the Cherenkov angle at which the pulse can be detected.  For a typical Cherenkov angle and beam scale width of \mbox{$\theta_{\rm c} \sim 0.4^\circ$} and \mbox{$\Delta\theta_{\rm max} \sim 0.02^\circ$} respectively (see \figref{fig:sim_beam}), this results in a geometric aperture of \mbox{$2 \pi R_{\rm J} \Delta R \, \Omega \sim 10^3$}~km$^2$\,sr, which is less than those of current detectors.

\begin{figure}
 \plotone{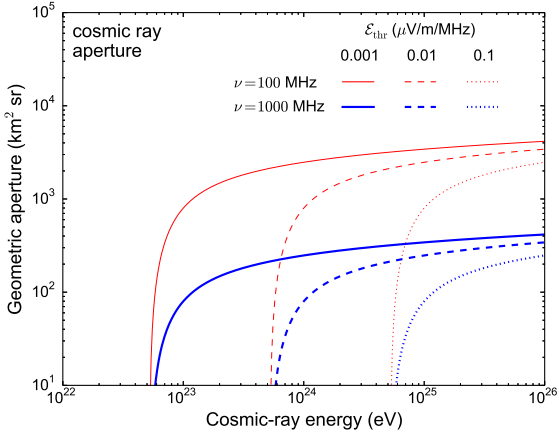}
 \caption{Geometric apertures of Earth-based radio antennas for the detection of cosmic rays interacting in the atmosphere of Jupiter.  The minimum detectable cosmic-ray energy is determined by the radio detection threshold $\mathcal{E}_{\rm thr}$, and the maximum geometric aperture is determined by the observing frequency $\nu$.}
 \label{fig:apertures}
\end{figure}

More generally, we can find the geometric aperture as
 \begin{equation}
  A(E; \mathcal{E}_{\rm thr}) = \int_{R_{\rm min}}^{R_{\rm max}}\! 2 \pi R_{\rm J} \, dR \,\, \Omega(\mathcal{E}_{\rm thr})
  \label{eqn:aperture}
 \end{equation}
where $\Omega(\mathcal{E}_{\rm thr})$ is the solid angle in which the pulse amplitude exceeds a detection threshold of $\mathcal{E}_{\rm thr}$, which describes the radio sensitivity of a coherent pulse detection experiment.  To calculate $\Omega(\mathcal{E}_{\rm thr})$, we first solve \eqnref{eqn:parameterisation} for $\Delta\theta$ with \mbox{$\mathcal{E} = \mathcal{E}_{\rm thr}$}, and then use this value as $\Delta\theta_{\rm max}$ when calculating $\Omega$ from \eqnref{eqn:solidang}.  We assume in \eqnref{eqn:parameterisation} that \mbox{$\rho = \rho_{\rm max}$} is evaluated at a column depth of \mbox{$X_{\rm max} = 1500$}~g/cm$^{2}$ along the shower axis, effectively assuming that the entire cascade occurs in a medium corresponding to the density at the shower maximum.  Across the altitude range $\Delta R$, these densities take values up to a maximum of \mbox{$3.6 \times 10^{-5}$}~g/cm$^{3}$, falling within the parameter range of our simulations in \secref{sec:emission}.  Some sample apertures for idealized Earth-based radio antennas are shown in \figref{fig:apertures}.  In \secref{sec:prospects}, we consider the prospects of detecting cosmic rays with some realistic experiments.

\section{Geometric aperture for neutrinos}
\label{sec:neutrinos}

\begin{figure}
 \plotone{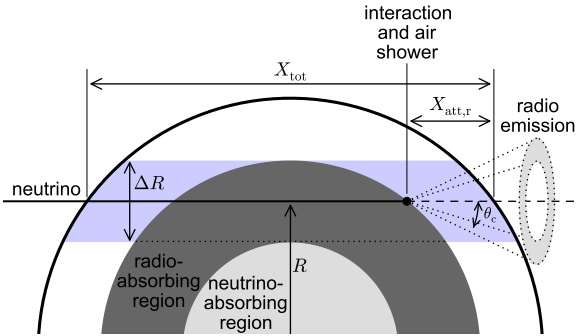}
 \caption{Interaction geometry for a detectable Jovian air shower initiated by a neutrino (cf.\ figure~\ref{fig:schematic}).  For the neutrino to be detectable, it must interact on its way out of the planet, with the remaining atmospheric column density along its path less than the radio attenuation length $X_{\rm att,r}$, so radio emission can escape from the resulting air shower (which has a length scale \mbox{$X_{\rm max} \ll X_{\rm att,r}$}).  This implies that the path of the neutrino should be sufficiently deep that the total column density \mbox{$X_{\rm tot} \gtrsim X_{\rm att,r}$}, so there is an appreciable chance that the neutrino will interact, but not so deep that $X_{\rm tot}$ exceeds the neutrino attenuation length $X_{{\rm att,}\nu}$, in which case it is likely to be absorbed deep within the planet.}
 \label{fig:schematic_nu}
\end{figure}

Neutrinos differ from cosmic rays in that they are able to propagate through a substantial depth of Jovian atmosphere before they interact and initiate an air shower.  However, ultra-high-energy neutrinos are still unable to propagate directly through the bulk of Jupiter, so detectable neutrinos are also constrained to trajectories that skim its atmosphere, albeit at greater depths than cosmic rays, as illustrated in figure~\ref{fig:schematic_nu}.  Since the radio attenuation length $X_{\rm att,r}$ is substantially smaller than the neutrino attenuation length $X_{{\rm att,}\nu}$, the majority of interacting neutrinos will produce air showers too deep in the atmosphere for the radio emission to escape.  As in \secref{sec:cosrays}, we will assume the radio emission from an air shower to escape if it propagates through a column density less than $X_{\rm att,r}$, and to be completely attenuated otherwise.

\begin{figure}
 \plotone{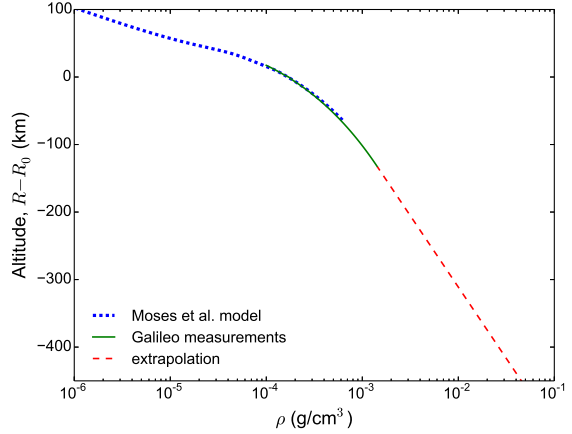}
 \caption{Density profiles for the atmosphere of Jupiter used in this work: from the model of \citet{moses2005}, from measurements by the Galileo probe \citep{seiff1998}, and an extrapolation of the latter with a constant scale height of 92~km.}
 \label{fig:atmoprofile}
\end{figure}

\begin{figure}
 \plotone{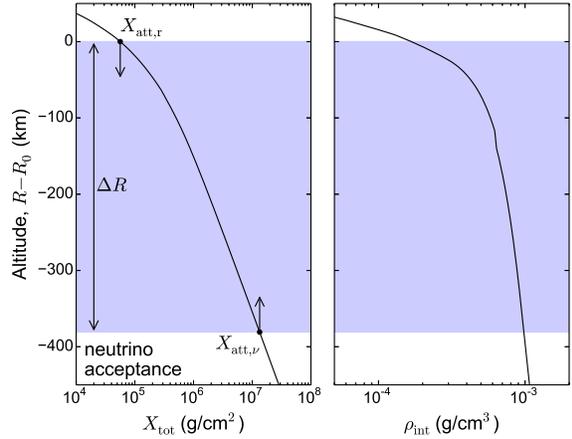}
 \caption{Range of altitudes (shaded) for the lowest point on the path of a detectable neutrino through the Jovian atmosphere, at radius $R$ as illustrated in \figref{fig:schematic_nu}.  As shown in the left panel, we require that the column density $X_{\rm tot}$ along the path of the neutrino is greater than the radio attenuation length $X_{\rm att,r}$ and lesser than the neutrino attenuation length $X_{{\rm att,}\nu}$.  This latter bound is slightly energy-dependent; the limit shown here is for \mbox{$E_\nu = 10^{23}$}~eV.  The right panel shows $\rho_{\rm int}$, the maximum density along the projected neutrino path within $X_{\rm att,r}$ of the exit point; this represents the point at which a detectable air shower is most likely to occur.}
 \label{fig:atmo_nu}
\end{figure}

\begin{figure}
 \plotone{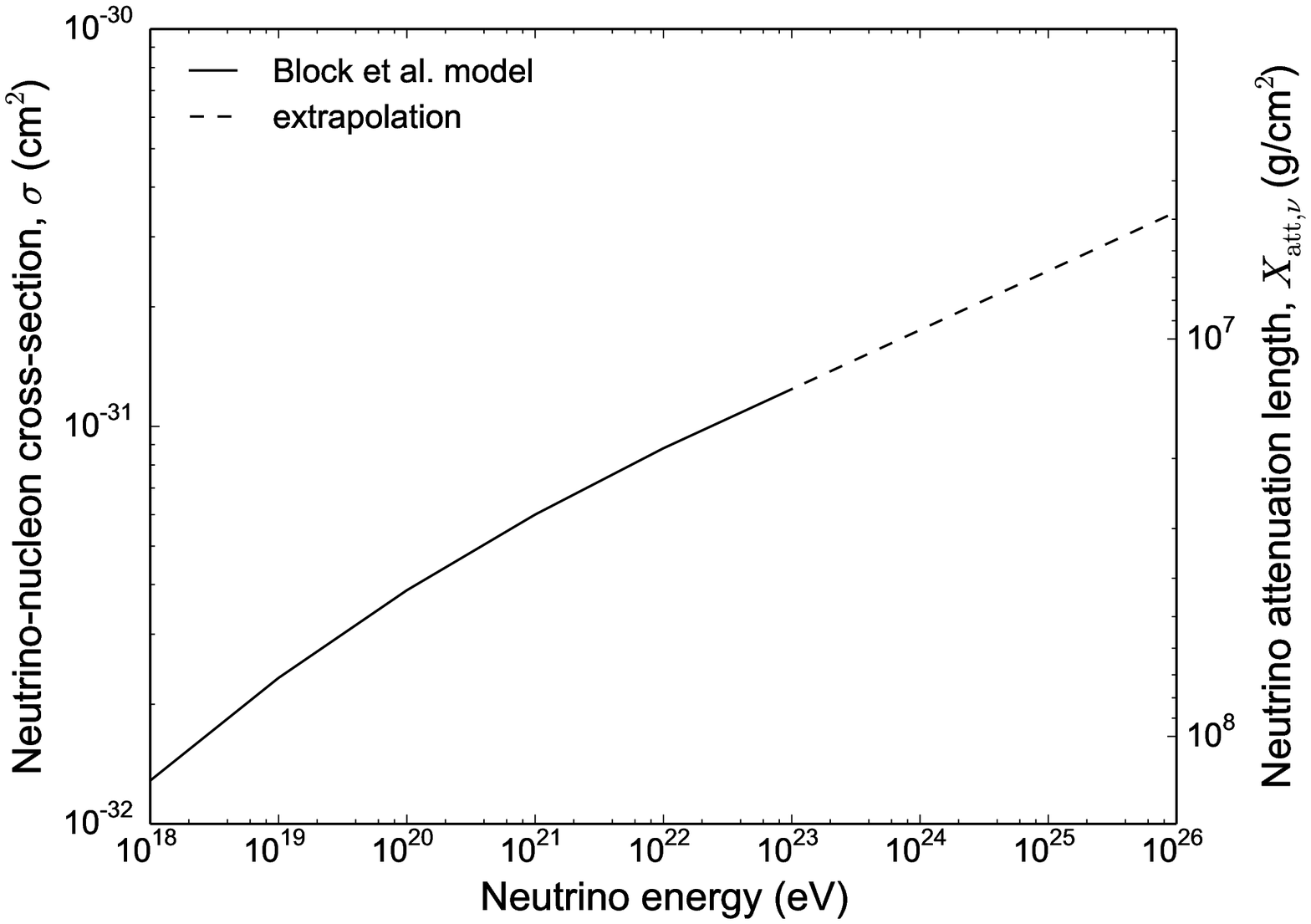}
 \caption{Total neutrino-nucleon interaction cross-section $\sigma$ from \citet{block2010} (solid) and a power-law extrapolation to higher energies (dashed).  The corresponding column density for neutrino attenuation is shown on the right axis, found as \mbox{$X_{{\rm att},\nu} = m_{\rm n} / \sigma$} where $m_{\rm n}$ is the nucleon mass.}
 \label{fig:nu_att}
\end{figure}

The deepest possible path for a detectable neutrino is determined by neutrino absorption in the lower atmosphere.  This regime lies beyond that described by the atmospheric model of \citet{moses2005} used in \secref{sec:cosrays}, so we extend it to lower altitudes using measurements from the Galileo mission \citep{seiff1998}, and extrapolate still further assuming a constant scale height (see \figref{fig:atmoprofile}).  As shown in \figref{fig:atmo_nu}, we require that the total column density $X_{\rm tot}$ along the path of the neutrino be less than the neutrino attenuation length $X_{{\rm att,}\nu}$, which we find from the neutrino interaction cross-sections modeled by \citet{block2010}, with a power-law extrapolation to higher energies (see \figref{fig:nu_att}).  This is an approximation, assuming that neutrinos experience no attenuation at less than a single attenuation length, and are completely attenuated beyond this threshold, as well as neglecting the recycling of interacting neutrinos to lower energies, which causes a factor \mbox{$\sim 1.4$} increase in the neutrino flux \citep{gayley2009}.  However, as in \secref{sec:cosrays}, modifying this assumption will have a relatively small effect on the minimum altitude $R_{\rm min}$ of the detection region.

For a neutrino to be detected, it must interact in the final radio attenuation length $X_{\rm att,r}$ along the total column density $X_{\rm tot}$, so the radio emission from the resulting air shower can escape.  If the neutrino follows a shallow path through the atmosphere, such that \mbox{$X_{\rm tot} < X_{\rm att,r}$}, it is proportionately less likely to interact.  As shown in \figref{fig:schematic_nu}, we neglect this reduced probability of interaction, taking the maximum altitude $R_{\rm max}$ of the detection region to be defined by the limit \mbox{$X_{\rm tot} = X_{\rm att,r}$}.  Neglecting those neutrinos that interact on shallower trajectories than this will cause us to underestimate the neutrino aperture, but this is a relatively small effect, both because these interactions are less common and because those air showers that do develop will do so in a less dense medium, and hence produce less radio emission.

% $\Delta R$ varies by \mbox{$\sim 30$}~km, or about 10\%, for each decade in energy.

Given these approximations, we can find the neutrino aperture similarly to \eqnref{eqn:aperture}, as
 \begin{equation}
  A(E_\nu; \mathcal{E}_{\rm thr}) = \frac{X_{\rm att,r}}{X_{{\rm att,}\nu}} \int_{R_{\rm min}}^{R_{\rm max}}\! 2 \pi R_{\rm J} \, dR \,\, \Omega(\mathcal{E}_{\rm thr})
  \label{eqn:aperture_nu}
 \end{equation}
where the initial ratio \mbox{$X_{\rm att,r} / X_{{\rm att,}\nu} \sim 10^{-2}$} represents the probability of the neutrino interacting within one radio attenuation length of the exit point.  In calculating $\Omega(\mathcal{E}_{\rm thr})$ here we assume that the shower develops in a medium with a density \mbox{$\rho = \rho_{\rm max}$}, the maximum atmospheric density along this final segment of the neutrino path, as shown in \figref{fig:atmo_nu}; note that these values are typically around $10^{-3}$~g/cm$^{3}$, so we are extrapolating outside the range of our simulations in \secref{sec:emission}.  In practice $\rho$ may take any value from $\rho_{\rm max}$ down to a minimum of near-zero density, for a neutrino that interacts in the outer reaches of the Jovian atmosphere, but the initial interaction is more likely to occur in a high-density region, hence our assumption.  We also assume that the shower energy is 20\% of the neutrino energy $E_\nu$, typical for the hadronic shower from a neutrino-nucleon interaction \citep{james2009b}.

\begin{figure}
 \plotone{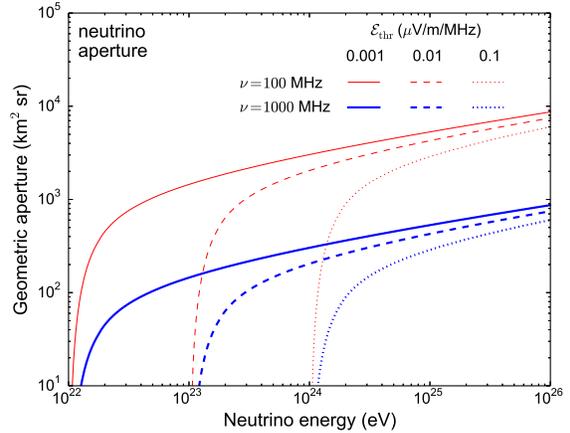}
 \caption{Geometric apertures of Earth-based radio antennas for the detection of neutrinos interacting in the atmosphere of Jupiter, for different radio detection thresholds $\mathcal{E}_{\rm thr}$ and observing frequencies $\nu$.  The apertures are generally larger, and the energy thresholds generally lower, than those shown for cosmic rays in \figref{fig:apertures}.}
 \label{fig:apertures_nu}
\end{figure}

In \figref{fig:apertures_nu} we show some sample neutrino apertures for the same idealized Earth-based radio antennas as for cosmic rays in \figref{fig:apertures}.  Compared to the case for cosmic rays, we might expect that the aperture would be reduced, due to the low probability that a neutrino will interact high enough in the atmosphere for the radio emission to escape, and that the energy threshold would be increased, as only a fraction of the original particle energy is manifested in the air shower.  However, air showers initiated by neutrinos typically begin deeper in the atmosphere and hence develop in a denser medium (see \figrefii{fig:atmo}{fig:atmo_nu}), which counteracts both of these effects: per \eqnref{eqn:parameterisation}, air showers developing in a denser medium will have stronger emission, allowing less-energetic showers to be detected; and the emission will be beamed at a larger Cherenkov angle, increasing the solid angle of the emission.  The net effect is that the aperture for detection of neutrinos is slightly larger than that for cosmic rays, and the energy threshold is slightly lower.  In \secref{sec:prospects}, we will examine how these apertures translate into prospects of detecting neutrinos with some realistic experiments.

\section{Detection prospects}
\label{sec:prospects}

To date, the experiment with the greatest sensitivity to coherent astronomical pulses is the LUNASKA Parkes experiment \citep{bray2014a}, which used the Parkes radio telescope to search for pulses from particle cascades in the lunar regolith.  The most sensitive future radio telescope currently being developed, Phase~1 of the Square Kilometre Array (SKA), will have substantially greater sensitivity in the same role \citep{bray2014b}.  If these telescopes were used instead to search for radio pulses from Jupiter, they would constitute experiments capable in principle of detecting Jovian air showers, albeit with some technical challenges such as compensating for dispersion of the pulse in the Jovian ionosphere.  Parameters for these experiments are listed in \tabref{tab:exps}.

\begin{deluxetable}{lcccc}
 \tablewidth{0pt}
 \tablecaption{Potential sensitivity of coherent pulse detection experiments \label{tab:exps}}
 \tablehead{
  \colhead{Experiment} &
  \colhead{Threshold} &
  \colhead{Frequency} &
  \colhead{Distance} &
  \colhead{Observing time}
  \\
  \colhead{} &
  \colhead{($\mathcal{E}_{\rm thr} / (\mu{\rm V/m/MHz})$)} &
  \colhead{($\nu / {\rm MHz}$)} &
  \colhead{($l / {\rm m}$)} &
  \colhead{($t_{\rm obs} / {\rm h}$)}
 }
 \startdata
  LUNASKA Parkes\tablenotemark{a} & 0.0047 &    1350 & $8 \times 10^{11}$ & \phn127.2 \\
%  NuMoon         & 0.1453 & \phn141 & $8 \times 10^{11}$ & \phn\phn46.7 \\
% Removed NuMoon because the sensitivity was substantially worse than these other experiments: the NuMoon limit in fig:cr_flux barely makes it into the top-right corner of the plot.
  SKA-lunar\tablenotemark{a}      & 0.0014 & \phn225 & $8 \times 10^{11}$ & 1000\phd\phn \\
%  Juno MWR       & 2.9\phn\phn\phn & \phn600 & $2 \times 10^{8}$\tablenotemark{a}$^{\phn}$ & 8760\tablenotemark{b}\phd\phn \\
% Replace Juno MWR with hypothetical satellite experiments.
  satellite (high altitude) & 0.23\phn\phn & \phn100 & $2.4 \times 10^7$ & 8760\phd\phn \\
  satellite (low altitude)  & 0.23\phn\phn & \phn100 & $1.2 \times 10^8$ & 8760\phd\phn \\
 \enddata
% \tablenotetext{a}{Minimum distance at which Jupiter fits within the antenna beam.}
% \tablenotetext{b}{Total nominal Jupiter-orbiting operation of the Juno mission.}
 \tablenotetext{a}{These experiments are/were not designed to detect Jovian air showers.  These parameters reflect their potential sensitivity if they had been so designed.}

 \tablerefs{\citep{bray2014a,bray2014b}}
\end{deluxetable}

\begin{figure}
 \plotone{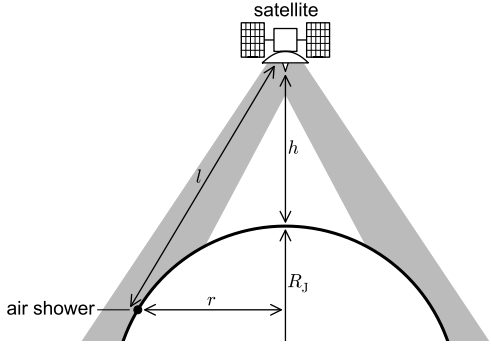}
 \caption{Geometry for radio air-shower detection with a Jupiter-orbiting satellite.  The antenna beams (shaded) must observe around the limb of the planet, so as to detect air showers in the atmosphere-skimming geometries shown in \figrefii{fig:schematic}{fig:schematic_nu}.  If the altitude $h$ of the satellite is decreased, there is a shorter distance $l$ to the air shower, allowing a less energetic shower to be detected.  However, the circumference $2 \pi r$ of the horizon viewed by the satellite is also decreased, leading to a reduced aperture for cosmic rays or neutrinos.}
 \label{fig:satellite}
\end{figure}

Another possibility is to search for radio pulses using an antenna on a Jupiter-orbiting satellite, which has the advantage of closer proximity to the source of the pulse.  To explore the potential of this approach, we consider a hypothetical satellite experiment with a radio collecting area of 100~m$^2$, and 100~MHz of bandwidth centered on an observing frequency of \mbox{$\nu = 100$}~MHz, close to the minimum-frequency limit imposed by the Jovian decametric radiation, to maximize the aperture.  The system noise at this frequency will be dominated by received radiation, from both Jupiter and the background sky; at this frequency the brightness temperature of the Galactic synchrotron background is \mbox{$\sim 1000$}~K \citep{thompson2001}, and the atmosphere of Jupiter is transparent down to an altitude with a similar temperature \citep{janssen2005}, so we take this value as the system temperature.  The noise level can then be calculated as $\mathcal{E}_{\rm rms} = 0.023$~$\mu$V/m/MHz \citep[][\eqn~7]{bray2016a}, and the threshold for a confident detection may reasonably be taken to be ten times this value.  The particle aperture will depend on the altitude of the satellite (see \figref{fig:satellite}), with a satellite closer to the planet being sensitive to less-energetic particles, but viewing a smaller volume of atmosphere.  We consider two scenarios with the satellite at altitudes of \mbox{$h = 4000$}~km, corresponding to the minimum perijove for the Juno probe \citep{janssen2014}, and \mbox{$h = R_{\rm J} = 69000$}~km, and assume that in either scenario the satellite could be operated for a period of one year.  Parameters for these experiments are also listed in \tabref{tab:exps}.

Note that this hypothetical satellite is optimistic compared to practical near-future instruments.  The collecting area and bandwidth are comparable to the existing space radio telescope RadioAstron \citep{kardashev2013}, but this instrument is in Earth orbit, not subject to the challenges of deployment to Jupiter or survival in the harsh Jovian environment, particularly its intense radiation belts.  The most sensitive radio instrument to be deployed this close to Jupiter is the microwave radiometer on the Juno probe \citep{janssen2014}, the largest antenna of which has an area of \mbox{$\sim 2$}~m$^2$ and a bandwidth of 21~MHz, substantially less than assumed here.  Furthermore, a single antenna beam does not have a sufficient field of view to view the entire horizon of Jupiter as required in this application; our hypothetical satellite would require multiple antennas or a phased array, with concomitant signal-processing requirements.

\begin{figure}
 \plotone{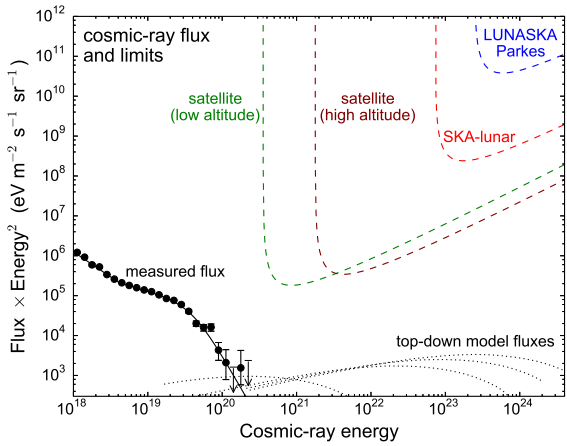}
 \caption{Ultra-high-energy cosmic-ray flux and potential limits from radio observations of Jupiter.  The points and solid line show the measured cosmic-ray flux and fit \citep{abraham2010}, while dotted lines show speculative cosmic-ray proton fluxes from decay of super-heavy dark matter \citep{aloisio2015}.  Dashed lines show potential limits from radio searches for Jovian air showers with the experiments described in the text.}
 \label{fig:cr_flux}
\end{figure}

We calculate the apertures of the Earth-based experiments for the detection of cosmic rays and neutrinos with \eqnrefii{eqn:aperture}{eqn:aperture_nu} respectively.  For the satellite experiments, we do the same, but replace the $2 \pi R_{\rm J}$ factor for the circumference of Jupiter with $2 \pi r$ for the circumference of the visible horizon (see \figref{fig:satellite}).  In both cases, we find the projected differential limits on the cosmic-ray or neutrino flux as
 \begin{equation}
  \frac{dF}{dE} < \frac{2.3}{E A(E) \, t_{\rm obs}} \label{eqn:limit}
 \end{equation}
where $t_{\rm obs}$ is the total observing time and the factor of 2.3 comes from the Poissonian distribution of the expected number of events for a 90\%-confidence limit.  These projected limits are shown for cosmic rays and neutrinos in \figrefii{fig:cr_flux}{fig:nu_flux} respectively.

The projected cosmic-ray limits, shown in \figref{fig:cr_flux}, do not extend down to sufficiently low energies to detect the known cosmic-ray flux.  Their energy range is suitable for testing predicted cosmic-ray fluxes from exotic top-down mechanisms, such as the decay of super-heavy dark matter \citep{aloisio2015}; models of this class are generally constrained by limits on the fluxes of ultra-high-energy neutrinos \citep{gorham2010} and photons \citep{abraham2009b}, but not entirely ruled out.  However, even these speculative fluxes are too low to be detected: the satellite experiments would need to be more sensitive by two orders of magnitude, while the experiments with Earth-based radio telescopes, LUNASKA Parkes and SKA-lunar, have even worse prospects.

\begin{figure}
 \plotone{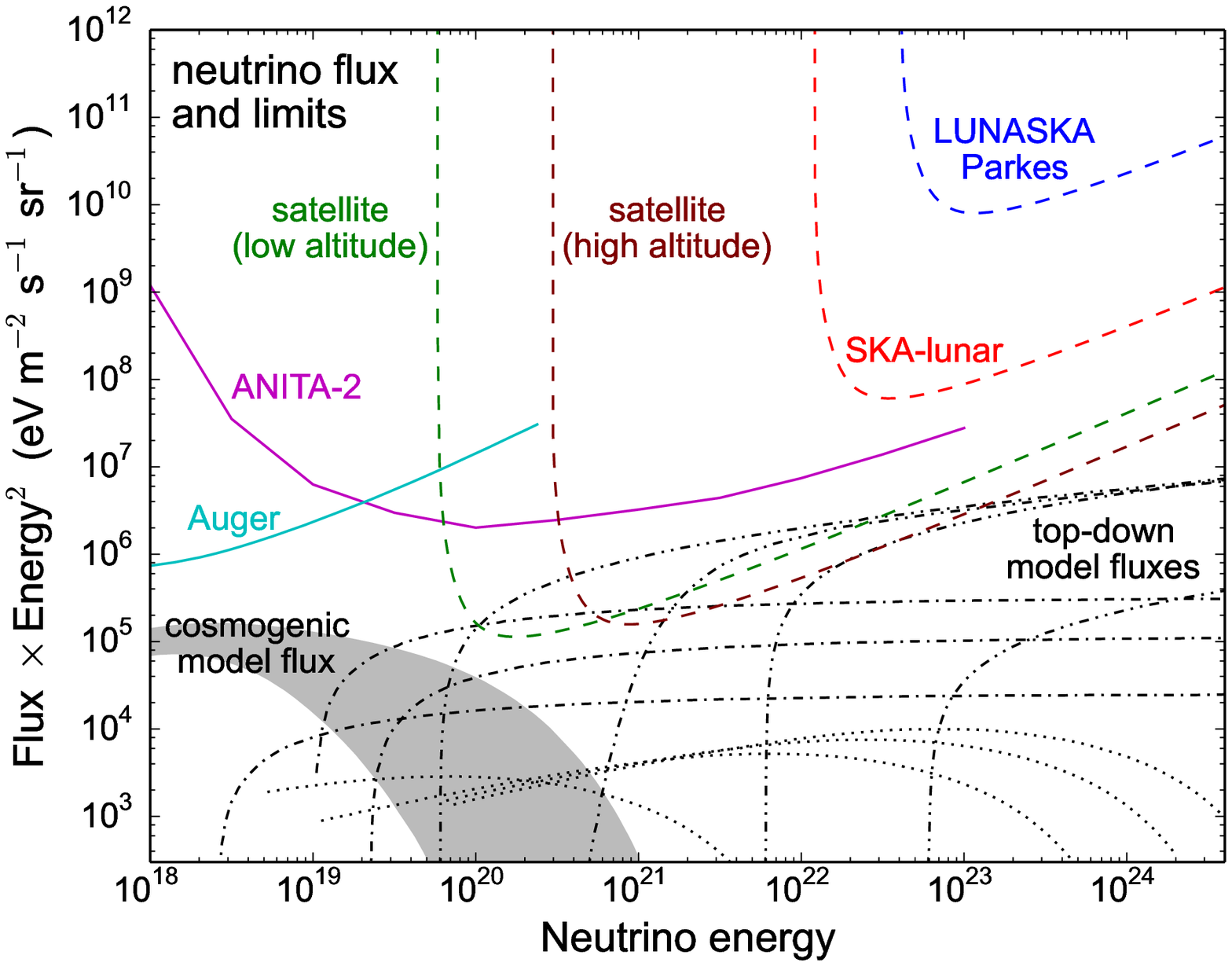}
 \caption{Models of the ultra-high-energy neutrino flux, existing limits, and potential limits from radio observations of Jupiter.  The shaded region shows a range of models for the expected cosmogenic neutrino flux \citep{kotera2010}, while more speculative models for neutrinos from various top-down mechanisms are shown from \citet{aloisio2015}, dotted; \citet{berezinsky2011}, dash-dotted; and \citet{lunardini2012}, dash-dot-dotted.  Solid lines show limits established by the ANITA experiment \citep{gorham2010} and the Pierre Auger Observatory \citep{aab2015} under the same definition as in \eqnref{eqn:limit}.  Dashed lines show potential limits from radio searches for Jovian air showers with the experiments described in the text.}
 \label{fig:nu_flux}
\end{figure}

The projected limits for neutrinos, shown in \figref{fig:nu_flux}, are slightly more promising.  The Earth-based experiments are still unable to detect the neutrino fluxes predicted from top-down mechanisms, but the satellite experiments are able to test some of the more optimistic models in this class.  The low-altitude satellite experiment has substantially better performance than its high-altitude equivalent, and is almost capable of detecting the cosmogenic neutrino flux expected from photopion interactions of propagating cosmic rays, regardless of the cosmic-ray origin model \citep{kotera2010}.  However, there are several projects under development, including ARA \citep{allison2012} and ARIANNA \citep{barwick2014}, that are likely to detect the cosmogenic neutrino flux in the near future, with substantially less risk and expense than a Jupiter-orbiting satellite.

\section{Discussion}
\label{sec:disc}

The prospect of utilizing Jupiter with its \mbox{$6 \times 10^{10}$}~km$^2$ of surface area as a detector for cosmic rays or neutrinos is an attractive one.  The only larger body in the solar system, the Sun, is a strong source of background radiation that makes it highly impractical to use in this role.  Consequently, no larger particle aperture than that of Jupiter will be available for the foreseeable future.

However, searches for radio emission from Jovian air showers are not a practical means of exploiting Jupiter as a particle detector.  We have found in this work that, with current or planned Earth-based radio telescopes, this technique is sensitive only to cosmic rays above an extremely high energy threshold, \mbox{$\sim 10^{23}$}~eV, and even then only to fluxes orders of magnitude higher than the most optimistic predictions in this energy range.  The situation with respect to neutrinos is similar.  A hypothetical highly-capable radio detector in close orbit around Jupiter might be able to test some of the more optimistic models of the neutrino flux, but this is a poor return for such a challenging experiment.
%  Observations at lower frequencies than those of the instruments considered here would increase the cosmic-ray aperture somewhat, but would be limited at frequencies below 40~MHz by the strong background of Jovian decametric radiation.

Our results here are not precise --- we have, for example, inaccuracy of a factor \mbox{$\sim 2$} in our parameterization of the radio emission from a shower --- but a more precise analysis is unlikely to be so much more optimistic than this work that it predicts this technique to be practical.  Our most serious approximation is probably our simplistic model of radio attenuation, which could be improved by implementing a radiative transfer model such as that of \citet{janssen2005}, but our assumptions err on the side of optimism, and a more sophisticated analysis is most likely to revise the sensitivity substantially downwards.

Our results conflict with those of \citet{rimmer2014}, who found that it was practical to detect radio emission from Jovian air showers with present-day instruments.  However, \citeauthor{rimmer2014}\ developed their own model of the radio emission in which it is represented as classical synchrotron radiation, which has proven to be a poor approximation to real air showers \citep[][\secname~3.1]{huege2016}.  As the ZHAireS code used in this work has been validated against observations of terrestrial air showers, we expect our results to be more reliable.

The major effect identified in this work is that, for the forward-beamed radio emission from a shower to escape Jupiter, the shower must be inclined almost horizontally, which causes it to develop in the upper atmosphere where the density is low.  In this environment, the radio emission is very narrowly beamed, which leads to a small cosmic-ray acceptance angle and hence a small geometric aperture.  This effect is less pronounced for neutrinos, which are able to pass more deeply through the planet and interact in the lower atmosphere on a trajectory that is inclined slightly upward.  Consequently, the aperture for neutrinos is larger than that for cosmic rays, though not sufficiently so to make detection practical.

We expect similar results to apply for the other gas giant planets in the solar system.  Each of them has a radius smaller than that of Jupiter, which allows a horizontally-inclined shower to develop at a lower altitude, but each also has a lower surface gravity and hence a larger atmospheric scale height, so the density at a given altitude will be reduced; combined, these effects may cause the typical density under which an air shower develops, and hence also the strength of its radio emission, to increase or decrease slightly.  The magnetic fields of the other gas giants are also weaker than the \mbox{$\sim 800$}~$\mu$T field of Jupiter \citep[e.g.\ Saturn, 21~$\mu$T;][]{davis1990}, which will reduce the strength of radio emission from showers in their atmospheres, though it will also ameliorate the shower suppression due to synchrotron losses.  Finally, all of the other gas giants are more distant than Jupiter, reducing the effectiveness of Earth-based instruments.

Our results also apply to other forms of radiation emitted by a Jovian air shower, provided that they are beamed at the Cherenkov angle and do not penetrate the lower atmosphere of Jupiter more efficiently than radio.  Optical Cherenkov radiation meets these conditions, and so is also excluded as a practical means of detecting Jovian air showers.

Our results do not exclude the possibility of practical detection of Jovian air showers through an emission mechanism that radiates isotropically, such as atmospheric fluorescence.  Orbital detection of atmospheric fluorescence from terrestrial air showers in the 330--400~nm band is being pursued by the JEM-EUSO project \citep{takahashi2009}, and the same technique could be applied with an imaging telescope in orbit around Jupiter.  The depth at which an air shower can be detected will be limited by Rayleigh scattering, for which an optical depth of unity for the JEM-EUSO band is reached in the Jovian atmosphere at a pressure of \mbox{$\sim 2000$}~mbar \citep[\fig~5.3]{west2004}, slightly deeper than the minimum altitude found in this work for radio detection of cosmic rays.  However, the JEM-EUSO band is tuned to the fluorescence spectrum of nitrogen, which is almost non-existent in the Jovian atmosphere.  Molecular hydrogen fluoresces primarily at wavelengths \mbox{$< 200$}~nm \citep{sternberg1989}, which are much more strongly affected by Rayleigh scattering; better results may be obtained with the helium line at 502~nm \citep{becker2010}, although helium is a minor constituent (\mbox{$\sim 25$}\% by mass) of the Jovian atmosphere.

\section{Summary}
\label{sec:conc}

We have investigated the potential for detecting radio emission from extensive air showers initiated in the atmosphere of Jupiter by ultra-high-energy cosmic rays or neutrinos.  We have developed models for the geometry under which such showers might be detected, simulated their development under representative Jovian conditions, and compared the strength of their radio emission to the sensitivity of current and planned Earth-based radio telescopes, and of a hypothetical radio detector in Jovian orbit.

We find that none of our considered experiments are likely to detect air showers initiated by cosmic rays, nor are Earth-based instruments likely to detect air showers initiated by neutrinos.  Under generous assumptions, a Jupiter-orbiting satellite may be able to detect some of the more optimistic predictions of the ultra-high-energy neutrino flux.

Our findings indicate that radio detection is unlikely to be a practical means to exploit the aperture of Jupiter as a particle detector.  Orbital fluorescence detection remains a possibility: its application to Jupiter seems to offer no compelling advantages, but may become worthwhile at some point in the future if the aperture provided by the Earth is fully exploited.

\acknowledgements
The authors would like to thank Jaime Alvarez-Mu\~{n}iz and Washington Carvalho Jr.\ for helpful discussions concerning ZHAireS, and an anonymous referee for several helpful comments regarding the content of the manuscript.  JDB acknowledges support from ERC-StG 307215 (LODESTONE). AN is supported by the DFG (research fellowship NE 2031/1-1). 

\bibliographystyle{apj}
\bibliography{all}

\end{document}